\documentclass[10pt,letterpaper]{article} % if you want global draft, use: [10pt,letterpaper,draft]

\usepackage[margin=1in]{geometry}
\usepackage[utf8]{inputenc}
\usepackage[T1]{fontenc}
\usepackage{lmodern}
\usepackage{textgreek}

\usepackage{graphicx}
\usepackage{caption}
\usepackage{wrapfig}
\usepackage{multirow}
\usepackage{amsmath}

\usepackage[backend=biber, maxnames=1, sorting=none]{biblatex}
\usepackage{hyperref}

\begin{document}

\begin{center}
{\Large
\textbf{Detecting Discontinuities in the Topology of Alzheimer's gene Co-expression}
}

\vspace{1em}

A. Samadzelkava\textsuperscript{1,2}

\vspace{0.5em}

\textbf{1} CUNY Graduate Center \\
\textbf{2} CCNY

\vspace{1em}

nsamadzelkava@gradcenter.cuny.edu
\end{center}

\begin{abstract}
Alzheimer's disease (AD) emerges from a complex interplay of molecular, cellular, and network-level disturbances that are not easily captured by traditional reductionist frameworks. Conventional analyses of gene expression often rely on thresholded correlation networks or clustering-based module detection, approaches that may obscure nonlinear structure and higher-order organization. Here, we introduce a comparative topological framework that makes use of topological data analysis (TDA) and Mapper algorithm to detect discontinuities - localized disruptions in the topology of gene co-expression space between healthy and AD brain tissue. Using gene expression data from 3 brain regions, we mapped how AD reshapes the global topology of gene–gene relationships. Discontinuity hotspots were identified via variability-based node scoring and subjected to GO Biological Process enrichment analysis.  This work illustrates the potential of TDA to uncover disease-relevant structure in high-dimensional transcriptomic data and motivates broader application of shape-based comparative methods in neurodegeneration research and other areas that benefit from comparative analysis.
\end{abstract}

\section*{Introduction}

Alzheimer's disease (AD) is a progressive neurodegenerative disorder characterized by cognitive decline, memory loss, and behavioral changes. It is the most common cause of dementia, accounting for 60–80\% of all dementia cases worldwide. The prevalence of AD is rapidly increasing, emphasizing the need for effective diagnostic and therapeutic strategies. Particularly in its late-onset form (LOAD), Alzheimer's disease epitomizes the complexity of human diseases, arising from a non-linear interplay between genetic variants, environmental factors, and molecular networks. Despite decades of intensive research and numerous identified risk factors, the causal mechanisms of LOAD remain elusive, and the quest for effective therapies has been fraught with challenges. Traditional focus on specific pathological features, such as amyloid-beta plaques and tau protein tangles, has not yielded the desired breakthroughs \cite{giuliaparoniUnderstandingAmyloidHypothesis2019, x.zhuangInterplayAccumulationAmyloidBeta2024, jiyoenkimTYK2RegulatesTau2024}. This highlights the need for a more holistic approach that embraces the multifaceted nature of the disease \cite{gabriellezunigaCenturyQuestionsRetrospective2015, rudyj.castellaniMolecularPathogenesisAlzheimers2009}.

Recent efforts to tackle the complexity of Alzheimer's disease include system-wide approaches aiming to pinpoint disruptions in the networks underlying the disease \cite{binzhangIntegratedSystemsApproach2013}. Gene coexpression networks, also known as covariance or correlation networks, present a successful translation of gene expression profiles into the language of network theory \cite{erice.schadtIntegrativeGenomicsApproach2005, horvathGeometricInterpretationGene2008}. Comparative analysis of the resultant networks between healthy and diseased conditions has been powerful in determining functional modules—groups of genes with similar expression patterns connected into functional complexes or pathways involved in regulatory or signaling circuits \cite{erice.schadtIntegrativeGenomicsApproach2005, juliepmerchantPredictiveNetworkAnalysis2022, lingfeiwangDetectionRegulatorGenes2016, metecivelekSystemsGeneticsApproaches2014, binzhangIntegratedSystemsApproach2013}. However, these traditional methods have limitations. Firstly, they often rely on arbitrary thresholding of the coexpression matrix to define network edges, which may exclude potentially important features of the network. Secondly, clustering algorithms used to identify functional modules are essentially point approximations, which may overlook significant features of the data, especially in the presence of complex, non-linear relationships. Furthermore, biological networks, including gene expression networks, exhibit complex structures at various scales, making them notoriously difficult to analyze using traditional clustering methods. The complexity of molecular landscape of LOAD demands analytical tools that can model the high-dimensional and complex relationships inherent in gene expression data \cite{metecivelekSystemsGeneticsApproaches2014}.

\subsection*{Topological Data Analysis}

Non-linear, data-driven methods are essential for uncovering hidden structures and patterns that linear, assumption-based approaches may miss \cite{yaraskafTopologicalDataAnalysis2022}. They can capture subtle variations and non-linear interactions critical for understanding disease mechanisms, providing a more accurate representation of the biological processes involved \cite{vandamGeneCoexpressionAnalysis2018}.

Topological Data Analysis (TDA) emerges as a promising candidate, offering a novel perspective that robustly addresses the limitations of traditional network-based approaches \cite{chazalIntroductionTopologicalData2021, carlssonTopologyData2009, lyuApplicationTopologicalData2025}. Grounded in topology, TDA extracts underlying structure of complex datasets, including high-dimensional gene expression data \cite{pataniaTopologicalGeneExpression2019}. Non-linear dimensionality reduction techniques, along with algorithms like Mapper, allow the extraction of complex structures and the overall shape of data without the need for simplification or assumptions about the data's shape. By focusing on qualitative topological features, TDA provides a framework for exploring the topology of gene coexpression networks, revealing patterns that are missed by traditional statistical methods \cite{Dey_2022, yaraskafTopologicalDataAnalysis2022}. This approach has the potential to uncover new insights into the dynamic interplay of genes and pathways involved in AD.

In this project we investigated misaligned genes associated with AD highlighted by by leveraging TDA's Mapper. We then extending the analysis to detect discontinuities that emerge in comparative analysis. Discontinuities that emerge during the analysis are caused by ``misaligned'' genes — not only those whose expression patterns deviate significantly from those in the healthy state but those who alter networks topology. Identifying such genes could provide valuable information on gene interactions and pinpoint intervention opportunities to restore the network to a healthy state. Ultimately, this topological approach aims to improve our understanding of the molecular mechanisms underlying Alzheimer's disease and contribute to the development of novel therapeutic strategies.

\section*{Results}

We applied the pipeline described below to analyze gene co-expression patterns across three brain regions—the dorsolateral prefrontal cortex (DLPFC), visual cortex (VC), and cerebellum—in healthy and Alzheimer's disease (AD) conditions. The primary objective was to highlight discontinuities in expression patterns and identify potentially “misplaced” genes whose connectivity diverges from healthy co-expression structure.

Our analysis is grounded in co-expression matrices, which capture the pairwise correlation of gene expression levels.  For each condition (healthy and AD) and brain region, we constructed a co-expression matrix.

The healthy co-expression network serves as a reference to define a notion of continuity in gene expression space. Mapper graphs constructed from this network were colored using a continuous color gradient that reflects local similarity in gene co-expression profiles. This approach highlights smooth transitions in expression—genes with similar co-expression are assigned similar colors.

To assess divergence in the disease state, we applied the color scheme derived from the healthy condition to the corresponding AD co-expression graph. Color discontinuities in the Alzheimer’s graph thus point to disrupted co-expression structure and altered gene relationships.

\renewcommand{\thesubsection}{\arabic{subsection}}
\subsection{Pipeline overview}
\subsubsection{Mapper Algorithm Overview}
The Mapper algorithm is a TDA technique that captures the “shape” of complex, high-dimensional datasets in a simplified graph \cite{chazalIntroductionTopologicalData2021}. First, a “lens” (or filter) function is applied to map each data point to a lower-dimensional scalar or vector. This range is then partitioned into overlapping intervals (covers). Within each interval, local subsets of data points are clustered, forming nodes in the Mapper graph. Nodes that share points from overlapping intervals are connected by edges, revealing how local clusters relate globally. This process preserves local groupings while highlighting overall structural features (i.e., topology) of the dataset \cite{carlssonTopologyData2009}.

\subsubsection{Color Continuity}
\begin{wrapfigure}{r}{0.6\textwidth}
  \centering
  \includegraphics[width=0.59\textwidth]{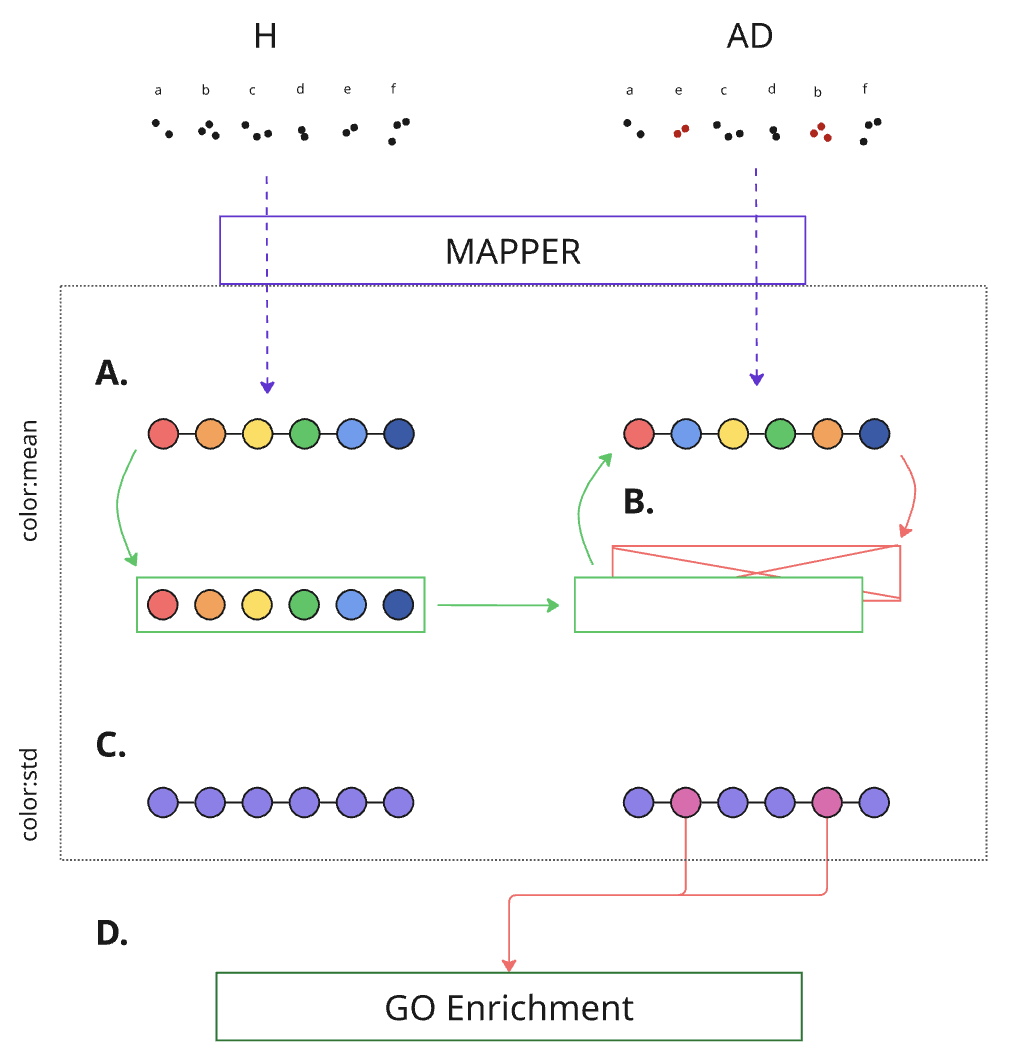}
  \caption{\textbf{A.} colors assigned to H. Mapper automatically assigns colors to AD but we discard those taking coloring of H as "ground truth".
\textbf{B.} We use colors H to color dataset AD highlighting discontinuities of AD
\textbf{C.} To highlight displaced genes in AD we apply "standard deviation" aggregation function to AD Mapper graph
\textbf{D.} Genes that fall into highly variable (pink) mapper nodes are extracted for GO enrichment analysis}
  \label{fig:mp}
\end{wrapfigure}
Mapper naturally facilitates smooth coloring of its resulting graph because it segments the data into overlapping bins along the chosen lens function. When assigning colors (for instance, by enumerating nodes and mapping them to a gradient), this adjacency ensures that similar or connected clusters receive closely related hues, leading to gradual shifts in color. As a result, regions of the graph that represent smoothly transitioning structures in the underlying data also display smooth color transitions, reflecting the continuous nature of the data’s topology (Figure~\ref{fig:mp}A).

Additionally, we rotated the color gradient by 90 degrees to investigate genes that might be displaced in a direction perpendicular to the initial color gradient, and therefore not highlighted by the original coloring axis (Figure \ref{fig:wrap1}A).

\subsubsection{Detecting Discontinuities}
Once a reference Mapper graph (from healthy controls) was assigned color scheme, we transferred those colors to the Mapper graph constructed from the AD dataset (containing the same genes but with different expression levels). If the AD dataset’s structure closely resembled the reference, the transferred colors appeared to flow smoothly across its nodes. However, when the AD dataset diverged from the reference, sharp color jumps—or discontinuities—emerged, indicating clusters or genes that shifted relative to their original placement. These discontinuities revealed where the AD topology deviated most from the baseline, highlighting potentially important structural differences (Figure~\ref{fig:mp}B).

To further quantify regions of altered connectivity, we applied standard deviation as an aggregation coloring function on the AD Mapper graph (Figure~\ref{fig:mp}C). Nodes with high color variability, where constituent genes derive from significantly different or conflicting color groups of the reference, stood out as hotspots of disrupted continuity. 

Building upon the identification of the top 20\% most variable Mapper nodes we compiled lists of genes from these nodes.Genes within these highly variable clusters were considered prime candidates for functional disruption linked to the disease.

\subsubsection{GO Enrichment Analysis}

For each region, we performed Gene Ontology (GO) enrichment analysis on the genes from the top-20\% variable clusters to determine which biological functions are most impacted under AD conditions. Enrichment was conducted using the Enrichr platform, focusing on GO-Biological Process terms. We applied a significance threshold of $p<0.05$ (after adjustment for multiple testing) and ranked GO terms by the combined score (CS) retaining only GO terms with $CS<30$. 
To benchmark our topological approach against conventional analyses, we also quantified differential gene expression and differential gene co-expression between control and AD samples (Figure\ref{fig:DLPFC_cnet}).

\begin{wrapfigure}{r}{0.6\textwidth}
  \centering
  \includegraphics[width=0.59\textwidth]{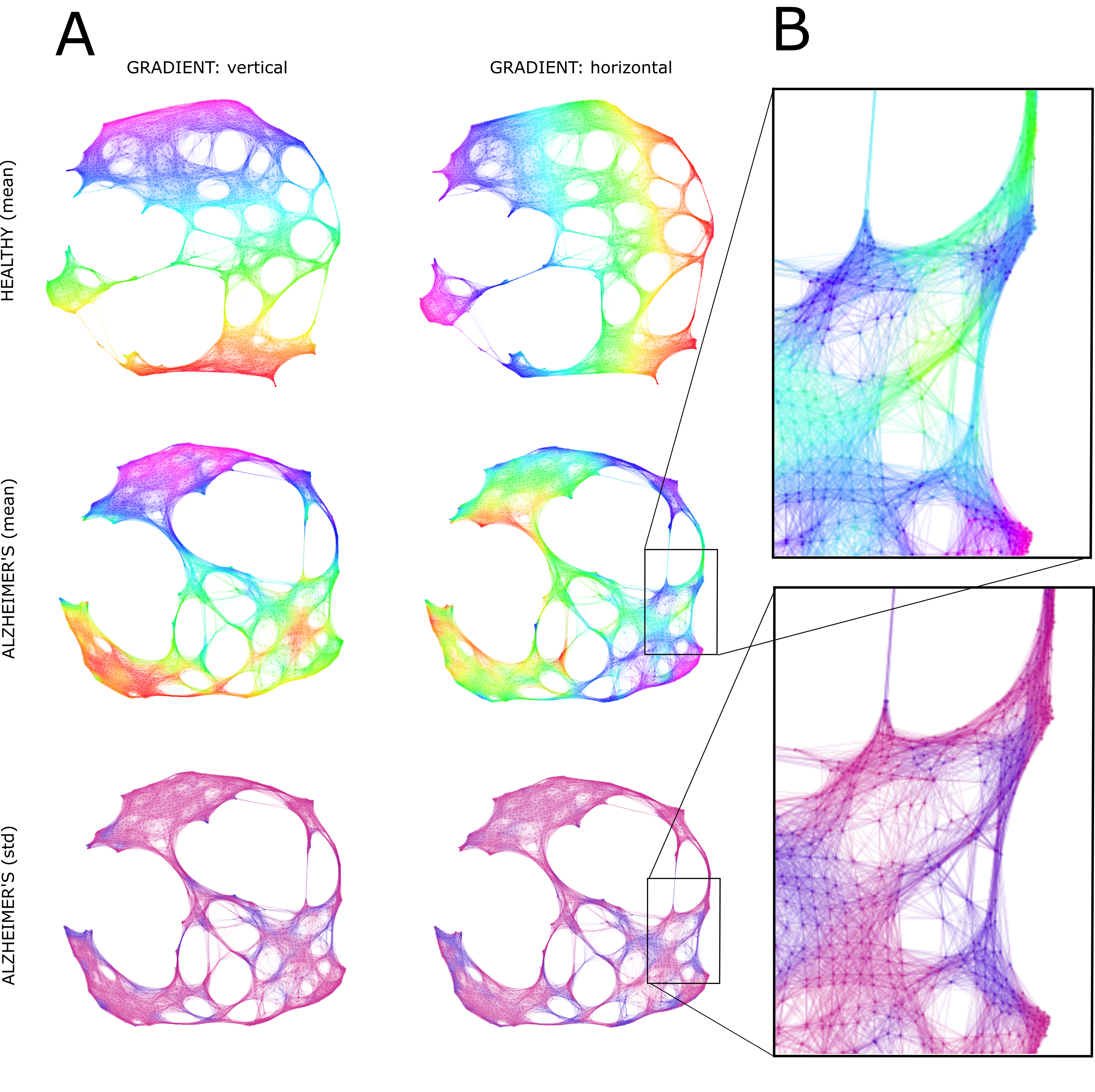}
  \caption{Mapper graph of DLPFC gene co-expression network (healthy reference coloring).  axes and Nodes representing gene clusters are colored according to the reference (healthy) smooth expression gradient. B. Discontinuities in the AD graph's colors indicate network perturbations in AD.}
  \label{fig:wrap1}
\end{wrapfigure}

\subsection{Findings}

\subsubsection{DLPFC}
The DLPFC is known to be one of the most affected brain regions in Alzheimer's disease, and our analysis supports this. Thirteen GO terms showed a combined score higher than 50, with the top terms associated with vascular smooth muscle cell development and differentiation, positive regulation of CD8-positive $\alpha\beta$ T cell differentiation, and protection from natural killer cell-mediated cytotoxicity. The enrichment of vascular-related pathways (e.g., vascular smooth muscle cell differentiation) aligns with previous research indicating vascular dysfunction as a critical factor in Alzheimer's pathology. Cerebrovascular abnormalities, including impaired vascular smooth muscle cell function, have been consistently associated with AD progression, emphasizing the role of compromised cerebral blood flow and vascular integrity in neurodegeneration \cite{Zlokovic2011, Iadecola2013}.

Additionally, immune response pathways, such as the regulation of T cell differentiation and protection from natural killer cell activity, underscore the emerging concept of immune dysregulation and inflammation in Alzheimer's pathology. Prior literature demonstrates significant T cell involvement and immune-mediated damage contributing to AD progression \cite{Gate2020, Bettcher2021}. Interestingly, several novel findings emerged from this analysis, including GO terms like positive regulation of isomerase activity and regulation of glial cell proliferation. These processes have less established roles in AD and may represent potential novel targets for therapeutic exploration or further mechanistic studies.

\clearpage
\begin{figure}[htbp]
  \centering
  \includegraphics[height=0.3\textheight]{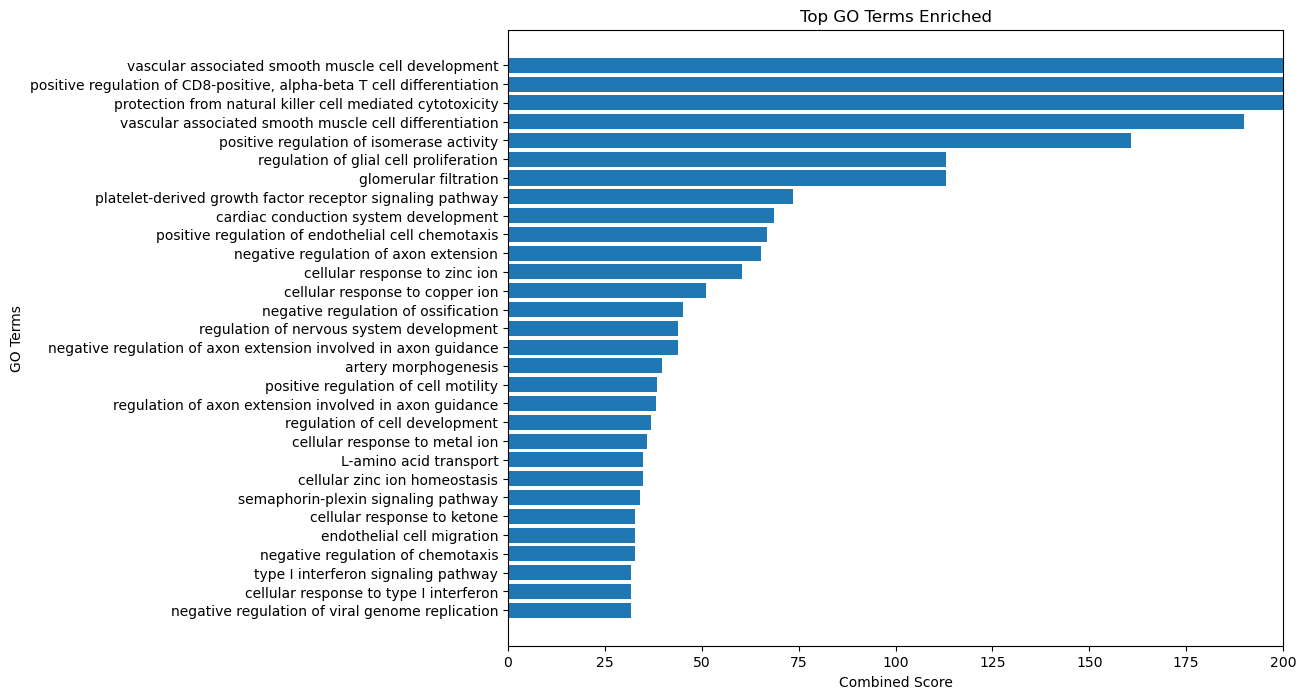}
  \caption{Top enriched GO terms in DLPFC sorted by combined score}
  \label{fig:DLPFC_bar_plot}
\end{figure}

\begin{figure}[htbp]
  \centering
  \includegraphics[height=0.3\textheight]{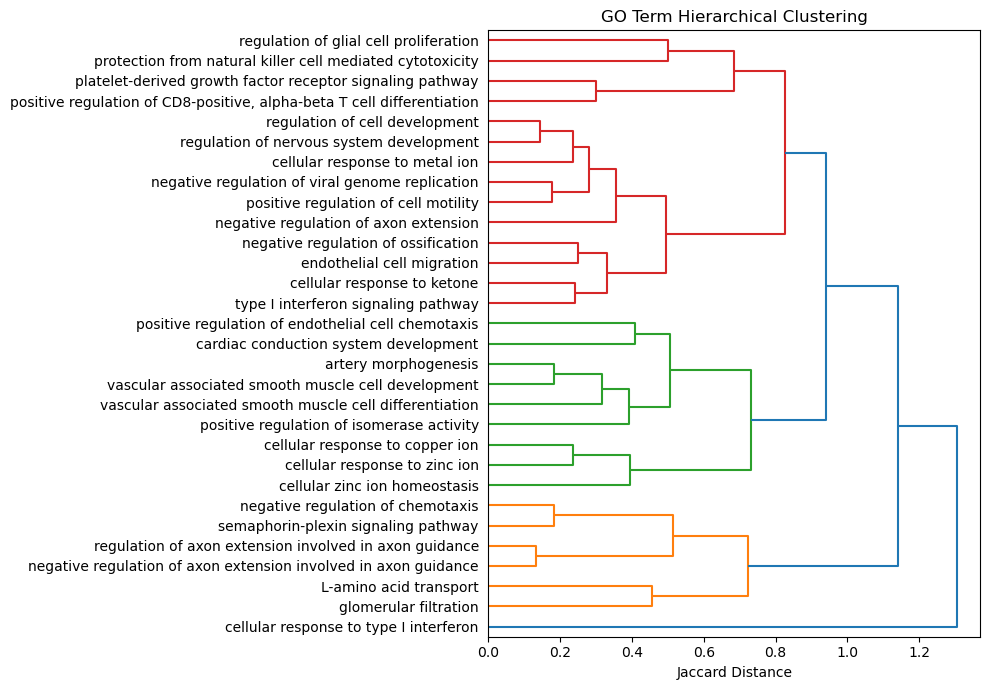}
  \caption{Hierarchical clustering of enriched GO terms in DLPFC.}
  \label{fig:DLPFC_dendro}
\end{figure}

Figure~\ref{fig:DLPFC_dendro} (DLPFC dendrogram) illustrates hierarchical clustering of enriched GO terms based on gene overlap, revealing four major clusters of biological processes in DLPFC. The largest cluster (red) predominantly includes immune and cellular regulation pathways (e.g., positive regulation of CD8-positive T cell differentiation, platelet-derived growth factor receptor signaling), consistent with literature emphasizing immune dysregulation in AD \cite{Gate2020, Bettcher2021}. Another cluster (green) groups vascular-associated processes such as vascular smooth muscle cell development and differentiation, aligning with known cerebrovascular dysfunction in AD \cite{Zlokovic2011, Iadecola2013}. An orange cluster highlights novel functional relationships, including regulation of axon extension, glomerular filtration, and cellular responses to interferon—pathways less well-established in AD and thus intriguing new avenues for investigation. The clear delineation between clusters underscores the complexity of AD mechanisms, suggesting both established and novel gene networks are at play in the DLPFC.

\begin{figure}[htbp]
  \centering
  \includegraphics[width=\textwidth]{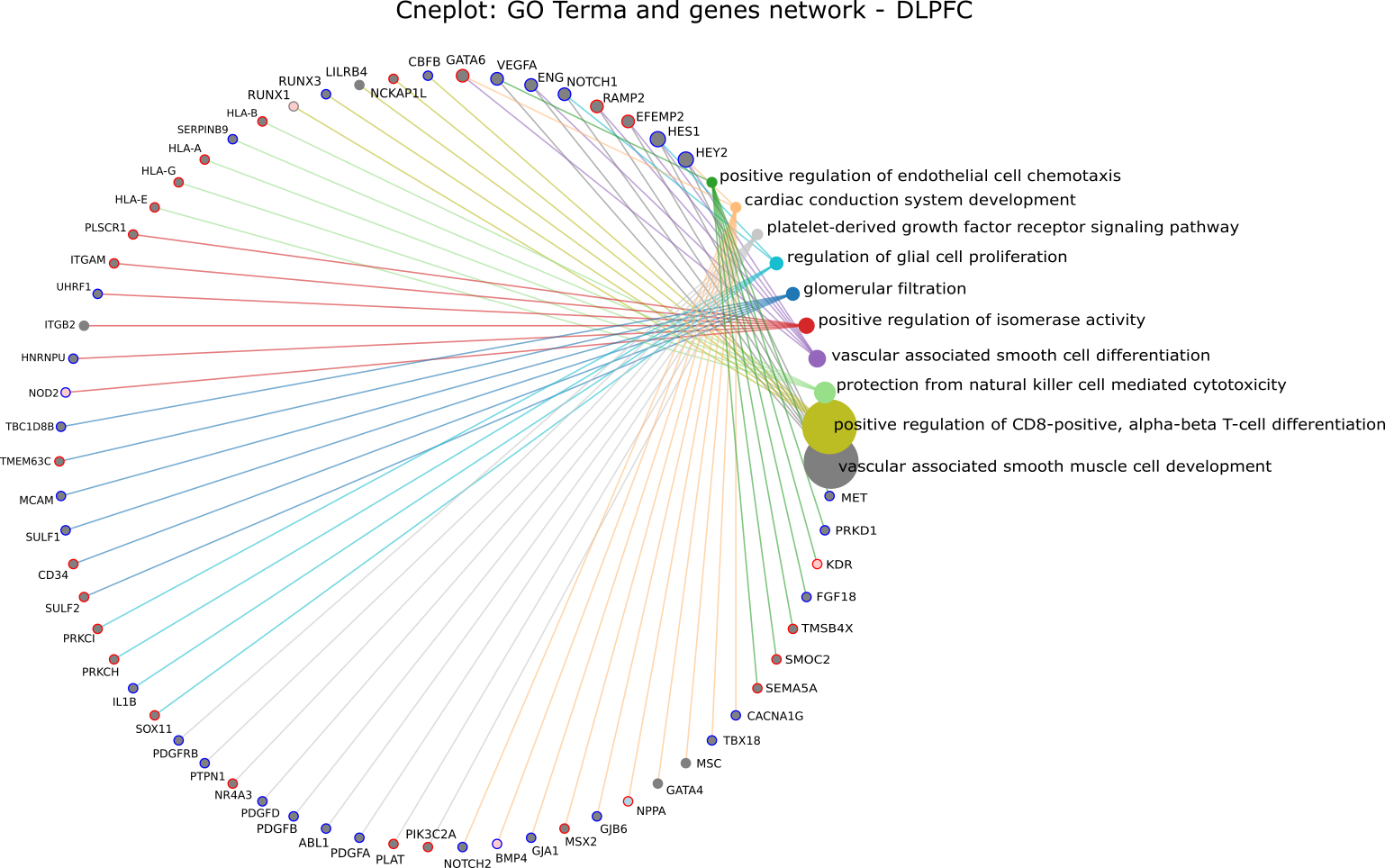}
  \caption{Top 10 GO term–gene network with enriched biological processes in DLPFC. GO terms scaled according to their combined enrichment scores. Genes are scaled by their in-degree (number of GO term associations). The interior color of each gene node reflects its differential expression status (up-regulated, down-regulated, or unchanged), whereas border color denotes the direction of its differential coexpression change between conditions.}
  \label{fig:DLPFC_cnet}
\end{figure}

The DLPFC cnetplot (Figure~\ref{fig:DLPFC_cnet}) maps connections between enriched GO terms and their associated genes, highlighting the diverse involvement of certain genes across multiple biological processes. Some genes (e.g., \textit{MET}, \textit{PRKD1}) appear only in single processes, indicating specialized roles. For example, \textit{MET} has been linked to neuronal survival pathways relevant to AD \cite{Wright2015}, whereas \textit{PRKD1}'s role in AD-related pathways is less documented, representing a novel area for exploration. In contrast, genes like \textit{HES1} span multiple GO terms, connecting vascular smooth muscle cell differentiation with glial cell proliferation—processes that intersect via Notch signaling \cite{Gridley2010, Lathia2008}. Such genes with broad involvement across pathways underscore the complexity of AD molecular networks and suggest that both specialized and multifunctional genes could be key therapeutic targets.

\subsubsection{Visual Cortex}

The visual cortex (VC) is traditionally considered less prominently affected in AD compared to regions like the DLPFC; neuropathological changes in VC often appear later and are less severe \cite{Braak1991, Thangavel2008}. Nonetheless, our analysis identified several significantly enriched GO terms in VC, suggesting subtle yet important biological disturbances. Eleven GO terms showed notable enrichment in AD, with “skeletal myofibril assembly,” “positive regulation of endothelial cell chemotaxis,” and “positive regulation of glial cell differentiation” among the most pronounced.

\begin{figure}[htbp]
  \centering
  \includegraphics[width=\textwidth]{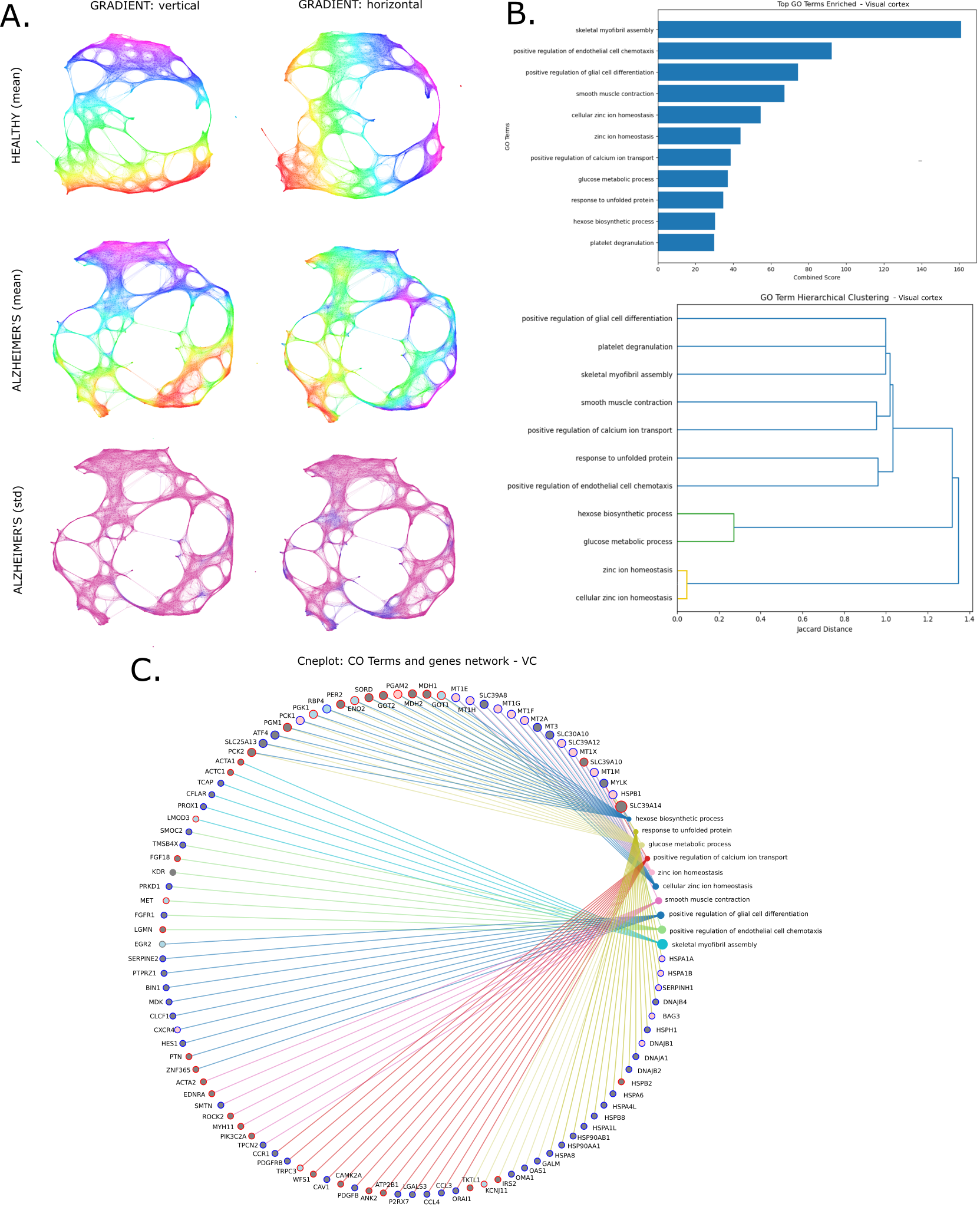}
  \caption{\textbf{A.}  Mapper graph of VC gene network (healthy coloring). Compared to DLPFC, the VC shows fewer disrupted regions (color discontinuities) in AD. \textbf{B.} Bar graph and hierarchical clustering of enriched GO terms in VC. \textbf{C.} Cneplot of VC (same as DLPFC) }
  \label{fig:VC_all}
\end{figure}
The enrichment of pathways related to endothelial chemotaxis and smooth muscle contraction supports evidence linking vascular dysfunction and cerebral hypoperfusion to AD pathology, even in less severely affected regions \cite{Zlokovic2011, Iadecola2013}. Additionally, GO terms like “cellular zinc ion homeostasis” and “response to unfolded protein” align with established molecular abnormalities in AD, particularly regarding protein misfolding stress and metal ion dysregulation \cite{Lovell1998, Hetz2017}. Interestingly, the significant enrichment of “skeletal myofibril assembly” and “platelet degranulation” in VC represents novel findings, as these processes have not been traditionally associated with AD pathology in this region, offering intriguing new targets for research.

The dendrogram for enriched GO terms in VC (Figure~\ref{fig:VC_all}B) reveals distinct clusters: the largest cluster (blue) groups terms related to vascular and stress-response pathways (e.g., endothelial cell chemotaxis, response to unfolded protein, smooth muscle contraction), consistent with vascular and proteostasis disturbances in AD \cite{Zlokovic2011, Hetz2017}. A smaller green cluster groups metabolic processes (hexose biosynthesis, glucose metabolism), aligning with known alterations in cerebral glucose metabolism—an AD hallmark \cite{Mosconi2005}. The orange cluster, marked by essentially zero Jaccard distance between terms, tightly pairs “zinc ion homeostasis” with “cellular zinc ion homeostasis,” highlighting metal ion dysregulation as a critical aspect in VC as well \cite{Lovell1998}. 

The VC cnetplot reveals that the gene \textit{SLC39A14} (ZIP14, a zinc transporter) is implicated in three key enriched processes: hexose biosynthetic process, cellular zinc ion homeostasis, and response to unfolded protein. This underscores \textit{SLC39A14}’s multifaceted role in the VC’s molecular landscape. As a member of the ZIP family of metal ion transporters (primarily Zn, but also Mn and Fe), \textit{SLC39A14} has established roles in cellular zinc regulation and metabolic signaling \cite{Fukada2018, Jenkitkasemwong2015}. Notably, recent preclinical AD models have demonstrated reduced ZIP14 levels in cortex under sporadic AD conditions—a deficit reversible with dietary zinc supplementation \cite{Rychlik2023}. Zinc dyshomeostasis is a recognized feature of AD pathology, correlated with A$\beta$ aggregation and impaired synaptic function \cite{Adlard2018, Sensi2018}. \textit{SLC39A14}’s association with hexose biosynthesis also suggests a metabolic dimension, resonating with its known involvement in glucose metabolism and insulin regulation in non-neuronal contexts \cite{Troche2016}. Meanwhile, its linkage to the unfolded protein response hints at a potential role in proteostasis under neurodegenerative stress \cite{Hetz2017}. Taken together, these connections bridge metal ion regulation, metabolic control, and protein-folding mechanisms in AD. \textit{SLC39A14} may serve as a nexus of vulnerability in cortical AD pathology, meriting deeper functional characterization.

\subsubsection{Cerebellum}
In contrast to DLPFC and VC, the cerebellum (CB) in our analysis did not show any GO Biological Process terms meeting our significance criteria after correction. This aligns with clinical and neuropathological observations that cerebellar involvement in AD is minimal compared to cortical regions. The lack of significant enrichment suggests that, within the sensitivity of our approach, the cerebellum's gene co-expression network remains relatively intact or unchanged in AD, or that any changes do not converge on common GO-defined pathways. While individual gene disruptions in CB may occur in AD, they might be too subtle or heterogeneous to yield a significant GO term enrichment. 

\subsection{Tau- and Amyloid-$\beta$–Relevant Processes Across DLPFC and Visual Cortex}

Across both DLPFC and VC, the enriched GO terms point to upstream pathways that modulate tau and amyloid-$\beta$ (A$\beta$) pathology. In DLPFC, several broad mechanistic themes emerge. Semaphorin–plexin signaling, PDGF$\rightarrow$c-Abl activity, metal-ion homeostasis, and type-I interferon responses all appear as upstream modulators of pathological tau, consistent with models in which kinase activation, Zn$^{2+}$/Cu$^{2+}$ imbalance, and inflammatory tone influence tau phosphorylation and aggregation \cite{eunheeahnMolecularMechanismsAlzheimers2025, jiyoenkimTYK2RegulatesTau2024, virendersinghZincPromotesLiquid2020, sophiesanfordTypeIInterferonResponse2023}. Notably, these pathways act without the need for increased \textit{MAPT} expression itself, aligning with contemporary views that tau dysfunction in AD arises from network-level stressors rather than direct \textit{MAPT} dysregulation.

A$\beta$-related processes in DLPFC map onto known modulators of A$\beta$ production, clearance, and toxicity. Lipid-handling and ApoE-associated pathways influence A$\beta$ deposition and removal, while immune and endo-lysosomal modules regulate plaque clearance and secretase activity \cite{christianbehl2024AmyloidcascadehypothesisStill2024, madialozuponeImpactApolipoproteinIsoforms2024, zezhonglvClearanceVamyloidSynapses2024}. Current evidence suggests a convergence between the two hallmark pathologies: A$\beta$ can amplify tau phosphorylation, and tau burden can exacerbate A$\beta$-driven synaptic dysfunction \cite{x.zhuangInterplayAccumulationAmyloidBeta2024}. The DLPFC GO-term profile reflects this interplay, highlighting pathways relevant to both arms of AD pathology.

In the visual cortex, enrichment patterns shift toward vascular, metabolic, and proteostasis pathways. Metal-ion regulation again appears, relevant to both tau phase behavior and A$\beta$ aggregation \cite{Abelein2023, Faller2009}. Calcium-handling terms relate to Ca$^{2+}$ dysregulation frameworks in which A$\beta$ exposure can trigger kinase cascades leading to tau hyperphosphorylation \cite{CalvoRodriguez2020}. Carbohydrate-metabolism terms implicate the O-GlcNAc axis, whose impairment promotes tau phosphorylation \cite{Liu2004, Liu2009}. ER stress and UPR-related processes correspond to PERK–eIF2$\alpha$ signaling changes that affect both BACE1 activity and neuronal tau burden \cite{Ajoolabady2022}. Vascular and platelet-associated pathways—tied to APP/secretase release from platelets and to endothelial A$\beta$ transport via receptors such as LRP1—connect VC enrichment to A$\beta$ handling and cerebrovascular contribution to AD \cite{Evin2012, Johnston2008, Inyushin2020, Petralla2024, Wang2021}. 

Taken together, the GO-term signatures from DLPFC and VC converge on broad biological themes—signaling stress, ion regulation, immune tone, vascular function, and proteostasis—that interface with both tau and A$\beta$ pathways, albeit with region-specific emphases.

\section*{Discussion}
This topological approach offers a new way to examine the structure of gene co-expression changes. By using Mapper to identify comparative discontinuities in data - that is, breaks or shifts in the topology of gene co-expression space relative to controls - we detect network changes that standard methods may miss. In the DLPFC, these shifts point to coordinated immune, vascular, and other cellular processes, consistent with the view that AD involves multiple interacting pathways.

There are, however, important limitations. The Mapper algorithm depends on several parameter choices (lens function, cover resolution, clustering threshold), and different settings can highlight different aspects of the dataset. While these parameters are more general and less restrictive than rigid assumptions about the shape of data, they might present challenges for data interpretation. It remains unclear which combinations of said parameters are most informative for biological interpretation, and this requires systematic evaluation. Our approach may also benefit from a percolation-like procedure to assess how stable these comparative discontinuities are across parameter ranges.

In summary, using a topological lens to track comparative discontinuities in TDA shows how localized genetic changes accumulate into broader network disturbances. This pipeline is extremely generalizable and can be applied to many forms of complex data - proteomic, epigenomic, imaging, longitudinal datasets or any data that requires comparison. Extending this work to other data types or time-series samples could clarify how these network disruptions arise and evolve over disease progression. Ultimately, topological tools in conjunction with discontinuity detection may help identify disruptions in complex biological systems.

\section*{Acknowledgments}
The author declares no competing interests.
I would like to thank A. Patania and C. Arenas-Mena for productive discussions. 

\section*{Data}
Data is publicly available at \url{https://www.ncbi.nlm.nih.gov/geo/query/acc.cgi?acc=GSE44772}

\section*{Code Availability}
Code can be found at \url{https://github.com/ayasamadzelkava/genes_Alzh_TDA}
\section*{Additional Resources}
Interactive map of genes and GO terms can be found at\\ \url{https://embed.kumu.io/ed76a53d20b9ad7cd4aac7dfd81103f0}

\section*{Materials and Methods}

\subsection*{Data Collection and Preprocessing}
\paragraph{Data Collection}
Our study utilized gene expression data from  GSE44772 dataset, obtained from the Gene Expression Omnibus (GEO) repository.    
The GSE44772 SuperSeries is composed of three SubSeries, each focusing on a different brain region: cerebellum, dorsolateral prefrontal cortex, and visual cortex. 
gene expression matrix contained \textit{G} = 39,302 probe-level measurements. Each individual contributed tissue from 3 brain regions. This SuperSeries provides multi-tissue gene expression profiles of the human brain \textit{(N = 230} individuals: 129 AD and 101 controls)), with a total of 690 samples across the three SubSeries. Gene expression profiling for these samples was conducted using the Rosetta/Merck Human 44K v1.1 microarray platform and contains  39,302 probe-level gene measurements..

\paragraph{Data Preprocessing}

Raw microarray expression values were examined for missing or infinite entries, which were filled using linear interpolation across samples. 
To mitigate confounding, expression values were adjusted for multiple sample-level covariates known to influence transcriptomic profiles: pH, age, RNA integrity number (RIN), gender, batch, post-mortem interval (PMI), tissue, and preservation method. 

\paragraph{Filtering genes.}
Microarray probes lacking a valid gene symbol (i.e., probes annotated as predicted or uncharacterized genes) were excluded. Only probes with an assigned \textit{Gene Symbol} were retained. 

Some genes were represented by multiple probes. To obtain a single expression value per gene, expression values from all probes mapping to the same gene symbol were averaged. 

\paragraph{Selection of the most variable genes.}
 To focus on genes exhibiting the strongest biological signal and reduce noise, we extracted the most variable subset of genes. Gene-wise variability was quantified using the interquartile range (IQR) across all samples. Genes were ranked by IQR, and the top one-third were retained. This yielded a final high-variability gene set comprising \textit{G} = 6,507 genes, which served as the input for downstream network-based analyses.

\paragraph{Co-expression Matrix Construction}
Using the preprocessed matrices, we computed Pearson correlation coefficients for each pair of genes, separately for AD and control groups in each brain region. This yielded a gene–gene co-expression matrix for every condition and region (dimensions $N \times N$ for $N$= 6,507 genes). These matrices form the input for subsequent topological analyses. 

\subsection*{Topological Data Analysis}

\subsubsection{Application of the Mapper Algorithm}

All Mapper computations were performed using the \texttt{python KeplerMapper} library. 

\paragraph{Dimensionality Reduction}

Dimensionality reduction was performed by utilizing a combination of Isomap and Uniform Manifold Approximation and Projection (UMAP).

\begin{itemize}
    \item \textbf{Isomap:} (n\_components=100) Isomap is a non-linear dimensionality reduction method that preserves the geodesic distances between all pairs of data points, effectively capturing the global manifold structure of the data. We set the number of components to 100 to retain sufficient variance and structural information from the original high-dimensional space.
    \item \textbf{UMAP:} (n\_components=2) After reducing the dimensions with Isomap, we further projected the data onto a two-dimensional space using UMAP.
\end{itemize}

The combination of Isomap and UMAP allows us to capture both the global and local structures of the gene expression data. Isomap preserves the overall manifold geometry, while UMAP maintains the local relationships between genes. 

\paragraph{Parameter selection}

We manually selected the covering parameters to maximize the topological complexity of the resulting Mapper graph—specifically aiming to reduce the number of disconnected components while increasing the number of loops or holes, which are indicative of nontrivial topological features. Parameters were chosen to ensure sufficient overlap between intervals, thereby promoting continuity in the graph, and to provide a sufficient number of intervals to capture fine-grained structural variations within the data by method of manual percolation and visual inspection.

Covering parameters were set as follows:

\begin{table}[h!]
\centering
\begin{tabular}{|l|c|c|c|}
\hline
\textbf{Brain Region} & \textbf{Parameter} & \textbf{Healthy} & \textbf{Alzheimer's} \\
\hline
\multirow{2}{*}{DLPFC} 
& Percentage overlap  & 0.72 & 0.72 \\
& Number of intervals     & 80   & 80 \\
\hline
\multirow{2}{*}{Cerebellum} 
& Percentage overlap  & 0.68 & 0.70 \\
& Number of intervals     & 80   & 80 \\
\hline
\multirow{2}{*}{Visual Cortex} 
& Percentage overlap  & 0.72 & 0.72 \\
& Number of intervals     & 80   & 80 \\
\hline
\end{tabular}
\caption{Mapper parameters used for each brain region under healthy and Alzheimer's conditions.}
\label{tab:mapper_params}
\end{table}

Density-Based Spatial Clustering of Applications with Noise (DBSCAN) algorithm was applied to cluster genes based on their expression profiles. DBSCAN identifies clusters as areas of high density separated by areas of low density, making it effective for detecting clusters of arbitrary shape and handling noise in the data.

The clustering step internally converts each pairwise correlation value $c$ into a distance using the standard transformation
\[
d = 1 - c,
\]
such that genes with higher correlation correspond to smaller distances. 

\paragraph{Gene Coloring}

Gene coloring is performed by KepplerMaper.
Each gene within the Mapper graph was assigned a color based on its DBSCAN cluster label derived from the healthy dataset. A distinct color was mapped to each unique cluster label, creating a visual representation of the clustering structure in the gene expression data. This color-coding allowed us to track how genes grouped in the healthy state are distributed in the Mapper graphs of both healthy and Alzheimer's datasets.

\paragraph{Node Coloring}

In the Mapper graph, nodes represent groups of genes. Since each node contains multiple genes, we employed two \textit{aggregation functions} to assign a single representative color to each node:
\begin{enumerate}
    \item Mean Color Value: (associated with DBSCAN cluster labels) of all genes within a node. This approach provides an average representation of the gene clusters present in the node.
    \item Standard Deviation of Colors:  to assess the variability of genes in the node.
\end{enumerate}

\subsubsection{Discontinuity Detection}

Discontinuity of AD dataset was highlighted by using Healthy dataset as a "continuous ground truth". 

Let $G$ be the set of genes and let $PC$ be the corresponding point cloud.  
Applying Mapper to $PC$ produces a Mapper graph $MG$ consisting of nodes
\[
MG\_node \subseteq G .
\]
Each node is assigned a numerical color value in the process of Mapper application
that results on a smooth color gradient on the graph. Each gene $g \in G$ inherits the color of the Mapper node it belongs to:
\[
g \in MG\_node \;\Longrightarrow\; \mathrm{color}(g)=MG\_node\_color(MG\_node).
\]

\vspace{1em}
To compare healthy and Alzheimer's data, we first compute $MG$ and $MG\_node\_color$ for the healthy point cloud.  
We then take the colors assigned to genes in the healthy case and transfer these gene-level colors to the Alzheimer’s Mapper graph $MG^{\mathrm{AD}}$.  
Thus, each gene in $MG^{\mathrm{AD}}$ receives the color associated with its label from the healthy graph.

By examining color transitions between adjacent nodes in $MG^{\mathrm{AD}}$, we identify discontinuities: abrupt changes in the transferred colors indicate regions where the Alzheimer’s structure diverges from the healthy organization.

\vspace{1em}
For each node $n\in MG^{\mathrm{AD}}$, we compute the standard deviation of the numerical color values of the genes it contains,
\[
\sigma(n) = \operatorname{StdDev}\{\mathrm{color}(g)\mid g\in n\},
\]
to quantify heterogeneity within the node.  
Nodes with the highest $\sigma(n)$ (top $20\%$ in each brain region: DLPFC, cerebellum, visual cortex) are selected as the most variable.  
Genes contained in these nodes correspond to regions exhibiting topological misalignment relative to the healthy state.

\paragraph{Rotation of the Coloring Gradient}

To ensure that topological discontinuities were not missed due to the orientation of the coloring gradient, we applied an additional relabeling step that effectively rotates the coloring axis. Mapper coloring based on a one–dimensional lens is sensitive only to variation along that specific axis; discontinuities aligned orthogonally will remain undetected. To address this, we reordered the cluster colors according to the vertical coordinate of each cluster's centroid in the two–dimensional lens space. For each cluster $c$, we computed its centroid position
\[
\bar{y}_{c} = \frac{1}{|S_c|}\sum_{i \in S_c} y_i,
\]
where $S_c$ denotes the set of points assigned to cluster $c$ and $y_i$ is the second coordinate of point $i$ in the projected space. Clusters were then sorted by $\bar{y}_{c}$, and new color labels were assigned in this top–to–bottom order. This procedure produces a coloring aligned with the orthogonal direction of the original gradient, allowing the Mapper analysis to detect structural discontinuities regardless of their orientation in the lens.

Extracted genes were used for consequent GO enrichment analysis.

\subsubsection{Mapper Graph Visualization}
Mapper graphs for figures (Figure\ref{fig:wrap1} and Figure\ref{fig:VC_all}A) were generated using Gephi. Sigmoid function was applied to STD color scheme to highlight regions with high standard deviation.

\subsection*{GO (Gene Ontology) Enrichment Analysis}

\subsubsection{Enrichr}

For the present study, we focused on Gene Ontology (GO) Biological Process (BP) annotations.
GO enrichment was performed with Enrichr (\url{https://maayanlab.cloud/Enrichr/}) via its Python API
(\texttt{gp.enrichr}), using the ``GO\_Biological\_Process\_2021'' gene set library and
\textit{Homo sapiens} as the organism. Enrichr combines results from multiple gene set libraries;
in our analysis we restricted attention to this GO BP 2021 library.

For each brain region, we thus obtained a table of enriched GO Biological Process terms with
associated statistics (including raw $p$-value, adjusted $p$-value, odds ratio, and combined
score). 
Adjusted p-value is computed using the Benjamini-Hochberg (BH) procedure to correct for multiple hypothesis testing and control the false discovery rate (FDR). Given a list of $m$ hypotheses with corresponding p-values $p_1, p_2, \dots, p_m$, sorted in ascending order, the BH-adjusted p-value for the $i$-th ranked term is:

\[
p_i^{\text{adj}} = \frac{p_i \cdot m}{i}
\]

where $i$ is the rank of the p-value in the sorted list and $m$ is the total number of terms tested. Terms with $p < 0.05$ (Fisher’s exact test, Benjamini--Hochberg correction for multiple testing) were considered significant. 
To quantify the prominence of each process, we
ranked GO terms by Enrichr’s combined score. The combined score is a composite metric that integrates the significance from the Fisher exact test (via the p-value) and the deviation from the expected rank (via a z-score). It is calculated as:

\[
\text{Combined Score} = \log(p) \cdot z
\]

where $p$ is the raw p-value from the enrichment test and $z$ is the z-score reflecting the deviation of the observed rank from the expected rank. 
A high combined score indicates that a GO term is both statistically significant (low $p$-value) and strongly overrepresented (high odds ratio) in the gene list. 
In addition, we applied a combined-score cutoff and retained only terms with combined score $\geq 30$.

\subsubsection{Hierarchical clustering/dendrogram}

To visualize functional similarity among enriched Gene Ontology (GO) terms, we constructed a hierarchical dendrogram based on shared gene content. First, we assembled a binary matrix \( M \in \{0, 1\}^{n \times g} \), where each row represents a GO term and each column represents a unique gene. Each entry \( M_{ij} = 1 \) if gene \( j \) is annotated to GO term \( i \), and \( 0 \) otherwise. 
We computed pairwise Jaccard distances between GO terms using:

\[
D_{J}(A, B) = 1 - \frac{|A \cap B|}{|A \cup B|}
\]

where \( A \) and \( B \) are the sets of genes associated with two GO terms to quantifies dissimilarity based on shared gene content. The resulting distance matrix was subjected to agglomerative hierarchical clustering using Ward’s linkage method, which minimizes the total within-cluster variance at each merging step.

\subsubsection{Cnetplot Construction: GO Term–Gene Bipartite Network}

To visualize the functional landscape of enriched biological processes, we constructed bipartite networks connecting GO terms and their associated genes using a custom circular network plot (cnetplot). From the list of enriched GO terms, we selected the top \( k \) terms by highest combined score (see Section~\textit{GO Enrichment Analysis}). Each GO term \( t_i \in T \) and gene \( g_j \in G \) was added as a node in a graph \( \mathcal{G} = (V, E) \), where \( V = T \cup G \), and an undirected edge \( (t_i, g_j) \in E \) was created if gene \( g_j \) was annotated to term \( t_i \).

\paragraph{Differential Expression Analysis}

Differential expression between control and Alzheimer’s disease (AD) samples was assessed at the gene level using a robust linear modeling framework. For each gene, expression values from all individuals were modeled using a robust linear regression of the form
\[
y_i = \beta_0 + \beta_1 \,\text{Condition}_i + \varepsilon_i,
\]
where $\text{Condition}_i$ indicates group membership (0 = control, 1 = AD), $\beta_1$ represents the estimated $\log_2$ fold change, and $\varepsilon_i$ captures residual noise. A p-value for differential expression was obtained for the coefficient $\beta_1$, and multiple hypothesis correction was performed using the Benjamini--Hochberg false discovery rate (FDR). Genes exceeding both a $\log_2$ fold-change threshold ($|\beta_1| > 1$) and an FDR-adjusted significance threshold ($q < 0.05$) were classified as differentially expressed and assigned a color label reflecting up-regulation (red), down-regulation (blue), or no significant change (gray) for visualization.

\paragraph{Differential Coexpression Analysis}

To characterize changes in gene--gene relationships between conditions, we computed separate Pearson correlation matrices for control and AD samples. A differential coexpression matrix was then obtained by subtracting control correlations from AD correlations,
\[
\Delta r_{ij} = r^{(\mathrm{AD})}_{ij} - r^{(\mathrm{Ctrl})}_{ij},
\]
for each gene pair $(i,j)$. For each gene, differential connectivity was summarized by computing (i) the average absolute change in correlation magnitude and (ii) the average signed change in correlation direction across all its gene--gene interactions. Genes with an average magnitude change exceeding a predefined threshold (0.5) were considered differentially coexpressed; positive shifts were assigned red labels, negative shifts blue labels, and non-significant changes gray. These labels were used to highlight genes showing substantial rewiring of coexpression structure between conditions.

%\bibliographystyle{unsrt}
%\bibliography{genes_TDA_citations}
\newpage
\printbibliography

@article{binzhangIntegratedSystemsApproach2013,
  title = {Integrated {{Systems Approach Identifies Genetic Nodes}} and {{Networks}} in {{Late-Onset Alzheimer}}’s {{Disease}}},
  author = {{Bin Zhang} and Zhang, Bin and {Chris Gaiteri} and Gaiteri, Chris and {Liviu‐Gabriel Bodea} and Bodea, Liviu-Gabriel and {Zhi Wang} and Wang, Zhi and {Joshua McElwee} and McElwee, Joshua J and {Alexei A. Podtelezhnikov} and Podtelezhnikov, Alexei A. and {Chunsheng Zhang} and Zhang, Chunsheng and {Tao Xie} and Xie, Tao and {Linh M. Tran} and Tran, Linh M. and Tran, Linh M and {Radu Dobrin} and Dobrin, Radu and {Eugene M. Fluder} and Fluder, Eugene M. and {Bruce E. Clurman} and Clurman, Bruce E. and {Stacey Melquist} and {Stacey Melquist} and Melquist, Stacey and Melquist, Stacey and {Manikandan Narayanan} and Narayanan, Manikandan and {Christine Suver} and Suver, Christine and {Hardik Shah} and Shah, Hardik and {Milind Mahajan} and Mahajan, Milind and {Tammy Gillis} and Gillis, Tammy and {Jayalakshmi Srinidhi Mysore} and Mysore, Jayalakshmi S. and {Marcy E. MacDonald} and MacDonald, Marcy E. and {John Lamb} and Lamb, John and Lamb, John and {David A. Bennett} and Bennett, David A. and {Cliona Molony} and Molony, Cliona and {David J. Stone} and Stone, David J. and {Vilmundur Gudnason} and Gudnason, Vilmundur and {Amanda Myers} and Myers, Amanda J. and {Eric E. Schadt} and Schadt, Eric E. and {Harald Neumann} and Neumann, Harald and Zhu, Jun and {Jun Zhu} and J, Zhu and Zhu, Jun and Zhu, Jun and {Valur Emilsson} and Emilsson, Valur},
  date = {2013-04-25},
  journaltitle = {Cell},
  volume = {153},
  number = {3},
  eprint = {23622250},
  eprinttype = {pmid},
  pages = {707--720},
  doi = {10.1016/j.cell.2013.03.030},
  pmcid = {3677161}
}

@article{carlssonTopologyData2009,
  title = {Topology and Data},
  author = {Carlsson, Gunnar},
  date = {2009-01-29},
  journaltitle = {Bulletin of the American Mathematical Society},
  shortjournal = {Bull. Amer. Math. Soc.},
  volume = {46},
  number = {2},
  pages = {255--308},
  issn = {0273-0979},
  doi = {10.1090/S0273-0979-09-01249-X},
  url = {http://www.ams.org/journal-getitem?pii=S0273-0979-09-01249-X},
  urldate = {2023-05-28},
  langid = {english}
}

@article{chazalIntroductionTopologicalData2021,
  title = {An {{Introduction}} to {{Topological Data Analysis}}: {{Fundamental}} and {{Practical Aspects}} for {{Data Scientists}}},
  shorttitle = {An {{Introduction}} to {{Topological Data Analysis}}},
  author = {Chazal, Frédéric and Michel, Bertrand},
  date = {2021-09-29},
  journaltitle = {Frontiers in Artificial Intelligence},
  shortjournal = {Front. Artif. Intell.},
  volume = {4},
  pages = {667963},
  issn = {2624-8212},
  doi = {10.3389/frai.2021.667963},
  url = {https://www.frontiersin.org/articles/10.3389/frai.2021.667963/full},
  urldate = {2025-08-07},
  langid = {english}
}

@article{christianbehl2024AmyloidcascadehypothesisStill2024,
  title = {In 2024, the Amyloid-Cascade-Hypothesis Still Remains a Working Hypothesis, No Less but Certainly No More},
  author = {{Christian Behl}},
  date = {2024},
  journaltitle = {Frontiers in Aging Neuroscience},
  eprint = {39295642},
  eprinttype = {pmid},
  doi = {10.3389/fnagi.2024.1459224},
  pmcid = {11408168}
}

@article{erice.schadtIntegrativeGenomicsApproach2005,
  title = {An Integrative Genomics Approach to Infer Causal Associations between Gene Expression and Disease},
  author = {{Eric E. Schadt} and Schadt, Eric E. and {John Lamb} and Lamb, John and {Xia Yang} and Yang, Xia and Zhu, Jun and {Jun Zhu} and J, Zhu and Zhu, Jun and Zhu, Jun and {Stephen W. Edwards} and Edwards, Steve and {Debraj GuhaThakurta} and {Debraj GuhaThakurta} and GuhaThakurta, Debraj and {Solveig K. Sieberts} and Sieberts, Solveig K. and {Stephanie A. Monks} and Monks, Stephanie A. and {Marc L. Reitman} and Reitman, Marc L. and {Chunsheng Zhang} and Zhang, Chunsheng and {Pek Yee Lum} and Lum, Pek Yee and {Amy Leonardson} and Leonardson, Amy and {Rolf Thieringer} and Thieringer, Rolf and {Joseph M. Metzger} and Metzger, Joseph M. and {Liming Yang} and Yang, Liming and {John C. Castle} and Castle, John C. and {Haoyuan Zhu} and Zhu, Haoyuan and {Shera Kash} and Kash, Shera F and {Thomas A. Drake} and Drake, Thomas A. and {Alan B. Sachs} and Sachs, Alan B. and {Aldons J. Lusis} and Lusis, Aldons J.},
  date = {2005-06-29},
  journaltitle = {Nature Genetics},
  volume = {37},
  number = {7},
  eprint = {15965475},
  eprinttype = {pmid},
  pages = {710--717},
  doi = {10.1038/ng1589},
  pmcid = {2841396}
}

@article{eunheeahnMolecularMechanismsAlzheimers2025,
  title = {Molecular {{Mechanisms}} of {{Alzheimer}}’s {{Disease Induced}} by {{Amyloid-β}} and {{Tau Phosphorylation Along}} with {{RhoA Activity}}: {{Perspective}} of {{RhoA}}/{{Rho-Associated Protein Kinase Inhibitors}} for {{Neuronal Therapy}}},
  author = {{Eun Hee Ahn} and {Jae-Bong Park}},
  date = {2025},
  journaltitle = {Cells},
  eprint = {39851517},
  eprinttype = {pmid},
  doi = {10.3390/cells14020089},
  pmcid = {11764136}
}

@article{gabriellezunigaCenturyQuestionsRetrospective2015,
  title = {A Century of Questions: {{Retrospective}} Study of the Controversy and Efficacy of {{Alzheimer}}'s Disease Models},
  author = {{Gabrielle Zuniga} and Zuniga, Gabrielle},
  date = {2015-05-01},
  doi = {10.15781/t2jd4pp3c}
}

@article{giuliaparoniUnderstandingAmyloidHypothesis2019,
  title = {Understanding the {{Amyloid Hypothesis}} in {{Alzheimer}}'s {{Disease}}},
  author = {{Giulia Paroni} and Paroni, Giulia and {Paola Bisceglia} and Bisceglia, Paola and {Davide Seripa} and Seripa, Davide},
  date = {2019-01-01},
  journaltitle = {Journal of Alzheimer's Disease},
  volume = {68},
  number = {2},
  eprint = {30883346},
  eprinttype = {pmid},
  pages = {493--510},
  doi = {10.3233/jad-180802}
}

@article{horvathGeometricInterpretationGene2008,
  title = {Geometric {{Interpretation}} of {{Gene Coexpression Network Analysis}}},
  author = {Horvath, Steve and Dong, Jun},
  editor = {Miyano, Satoru},
  date = {2008-08-15},
  journaltitle = {PLoS Computational Biology},
  shortjournal = {PLoS Comput Biol},
  volume = {4},
  number = {8},
  pages = {e1000117},
  issn = {1553-7358},
  doi = {10.1371/journal.pcbi.1000117},
  url = {https://dx.plos.org/10.1371/journal.pcbi.1000117},
  urldate = {2024-07-12},
  langid = {english}
}

@article{jiyoenkimTYK2RegulatesTau2024,
  title = {{{TYK2}} Regulates Tau Levels, Phosphorylation and Aggregation in a Tauopathy Mouse Model},
  author = {{Jiyoen Kim} and {Bakhos Tadros} and {Yan Hong Wei Liang} and {Youngdoo Kim} and {Cristian A. Lasagna-Reeves} and {Jun Young Sonn} and {D. Chung} and {Bradley T. Hyman} and {David M Holtzman} and {H. Zoghbi}},
  date = {2024},
  journaltitle = {Nature Neuroscience},
  eprint = {39528671},
  eprinttype = {pmid},
  doi = {10.1038/s41593-024-01777-2},
  pmcid = {11614740}
}

@article{juliepmerchantPredictiveNetworkAnalysis2022,
  title = {Predictive {{Network Analysis Identifies}}{{{\emph{JMJD6}}}}and {{Other Novel Key Drivers}} in {{Alzheimer}}’s {{Disease}}},
  author = {{Julie P Merchant} and {Julie P. Merchant} and {Kuixi Zhu} and {Kuixi Zhu} and Henrion, Marc and {Marc Henrion} and {Syed S.A Zaidi} and {Syed Shujaat Ali Zaidi} and {Lau Branden} and {Branden Lau} and {Sara Moein} and {Sara Moein} and {Melissa L Alamprese} and {Melissa Alamprese} and {Richard V Pearse} and {Richard V. Pearse} and {David A Bennett} and {David A. Bennett} and {Nilufer Ertekin-Taner} and {Nilüfer Ertekin-Taner} and {Tracy L Young-Pearse} and {Tracy L. Young‐Pearse} and {Rui Chang} and {Rui Chang}},
  date = {2022-10-22},
  journaltitle = {Cold Spring Harbor Laboratory - bioRxiv},
  doi = {10.1101/2022.10.19.512949}
}

@article{lingfeiwangDetectionRegulatorGenes2016,
  title = {Detection of {{Regulator Genes}} and {{eQTLs}} in {{Gene Networks}}},
  author = {{Lingfei Wang} and Wang, Lingfei and Wang, Lingfei and {Tom Michoel} and Michoel, Tom},
  date = {2016-01-01},
  journaltitle = {Springer International Publishing},
  pages = {1--23},
  doi = {10.1007/978-3-319-43335-6_1}
}

@article{lyuApplicationTopologicalData2025,
  title = {Application of {{Topological Data Analysis}} in {{Complex Network Structure Identification}}},
  author = {Lyu, Zhengqian},
  date = {2025},
  volume = {136},
  langid = {english}
}

@article{madialozuponeImpactApolipoproteinIsoforms2024,
  title = {Impact of Apolipoprotein {{E}} Isoforms on Sporadic {{Alzheimer}}’s Disease},
  author = {{Madia Lozupone} and {Francesco Panza}},
  date = {2024-01-01},
  journaltitle = {Neural Regeneration Research},
  volume = {19},
  number = {1},
  eprint = {37488848},
  eprinttype = {pmid},
  pages = {80--83},
  doi = {10.4103/1673-5374.375316},
  pmcid = {10479857}
}

@article{metecivelekSystemsGeneticsApproaches2014,
  title = {Systems Genetics Approaches to Understand Complex Traits},
  author = {{Mete Civelek} and Civelek, Mete and {Aldons J. Lusis} and Lusis, Aldons J.},
  date = {2014-01-01},
  journaltitle = {Nature Reviews Genetics},
  volume = {15},
  number = {1},
  eprint = {24296534},
  eprinttype = {pmid},
  pages = {34--48},
  doi = {10.1038/nrg3575},
  pmcid = {3934510}
}

@article{pataniaTopologicalGeneExpression2019,
  title = {Topological Gene Expression Networks Recapitulate Brain Anatomy and Function},
  author = {Patania, Alice and Selvaggi, Pierluigi and Veronese, Mattia and Dipasquale, Ottavia and Expert, Paul and Petri, Giovanni},
  date = {2019-01},
  journaltitle = {Network Neuroscience},
  shortjournal = {Network Neuroscience},
  volume = {3},
  number = {3},
  pages = {744--762},
  issn = {2472-1751},
  doi = {10.1162/netn_a_00094},
  url = {https://direct.mit.edu/netn/article/3/3/744-762/2179},
  urldate = {2024-08-27},
  langid = {english}
}

@article{rudyj.castellaniMolecularPathogenesisAlzheimers2009,
  title = {Molecular Pathogenesis of {{Alzheimer}}'s Disease: Reductionist versus Expansionist Approaches.},
  author = {{Rudy J. Castellani} and Castellani, Rudy J. and {Xiongwei Zhu} and Zhu, Xiongwei and {Hyoung Gon Lee} and Lee, Hyoung Gon and {Mark A. Smith} and Smith, Mark A. and {George Perry} and Perry, George},
  date = {2009-03-26},
  journaltitle = {International Journal of Molecular Sciences},
  volume = {10},
  number = {3},
  eprint = {19399255},
  eprinttype = {pmid},
  pages = {1386--1406},
  doi = {10.3390/ijms10031386},
  pmcid = {2672036}
}

@article{sophiesanfordTypeIInterferonResponse2023,
  title = {The type‐{{I}} Interferon Response Potentiates Seeded Tau Aggregation and Exacerbates Tau Pathology},
  author = {{Sophie Sanford} and {Lauren V. C. Miller} and {Marina Vaysburd} and {Sophie Keeling} and {Benjamin J. Tuck} and {Jason D. Clark} and {Michal Neumann} and {Leo C. James} and {William A. McEwan}},
  date = {2023-10-17},
  journaltitle = {Alzheimer's \& Dementia},
  eprint = {37849026},
  eprinttype = {pmid},
  doi = {10.1002/alz.13493},
  pmcid = {10916982}
}

@article{vandamGeneCoexpressionAnalysis2018,
  title = {Gene Co-Expression Analysis for Functional Classification and Gene–Disease Predictions},
  author = {family=Dam, given=Sipko, prefix=van, useprefix=true and Võsa, Urmo and family=Graaf, given=Adriaan, prefix=van der, useprefix=true and Franke, Lude and family=Magalhães, given=João Pedro, prefix=de, useprefix=true},
  date = {2018-07-20},
  journaltitle = {Briefings in Bioinformatics},
  shortjournal = {Briefings in Bioinformatics},
  volume = {19},
  number = {4},
  pages = {575--592},
  issn = {1477-4054},
  doi = {10.1093/bib/bbw139},
  url = {https://doi.org/10.1093/bib/bbw139},
  urldate = {2024-05-28}
}

@article{virendersinghZincPromotesLiquid2020,
  title = {Zinc Promotes Liquid–Liquid Phase Separation of Tau Protein},
  author = {{Virender Singh} and Singh, Virender and {Ling Xu} and Xu, Ling and {Solomiia Boyko} and Boyko, Solomiia and {Krystyna Surewicz} and Surewicz, Krystyna and {Witold K. Surewicz} and Surewicz, Witold K.},
  date = {2020-05-01},
  journaltitle = {Journal of Biological Chemistry},
  volume = {295},
  number = {18},
  eprint = {32229582},
  eprinttype = {pmid},
  pages = {5850--5856},
  doi = {10.1074/jbc.ac120.013166}
}

@article{x.zhuangInterplayAccumulationAmyloidBeta2024,
  title = {The {{Interplay Between Accumulation}} of {{Amyloid-Beta}} and {{Tau Proteins}}, {{PANoptosis}}, and {{Inflammation}} in {{Alzheimer}}'s {{Disease}}.},
  author = {{X. Zhuang} and {Jie Lin} and {Yamin Song} and {Ru Ban} and {Xin Zhao} and {Zhangyong Xia} and {Zheng Wang} and {Guifeng Zhang}},
  date = {2024},
  journaltitle = {Neuromolecular medicine},
  eprint = {39751702},
  eprinttype = {pmid},
  doi = {10.1007/s12017-024-08815-z}
}

@article{yaraskafTopologicalDataAnalysis2022,
  title = {Topological {{Data Analysis}} in {{Biomedicine}}: {{A Review}}},
  author = {{Yara Skaf} and {Yara Skaf} and {Reinhard Laubenbacher} and {Reinhard Laubenbacher}},
  date = {2022-05-01},
  journaltitle = {Journal of Biomedical Informatics},
  eprint = {35508272},
  eprinttype = {pmid},
  pages = {104082--104082},
  doi = {10.1016/j.jbi.2022.104082}
}

@article{zezhonglvClearanceVamyloidSynapses2024,
  title = {Clearance of β-Amyloid and Synapses by the Optogenetic Depolarization of Microglia Is Complement Selective},
  author = {{Zezhong Lv} and {Lixi Chen} and {Ping Chen} and {Huipai Peng} and {Yi Rong} and {Wei Hong} and {Qiang Zhou} and {Nan Li} and {Boxing Li} and {R. Paolicelli} and {Yang Zhan}},
  date = {2024},
  journaltitle = {Neuron},
  eprint = {38295790},
  eprinttype = {pmid},
  doi = {10.1016/j.neuron.2023.12.003}
}

@article{Dey_2022,
	author = {Tamal K. Dey and Tamal K Dey and Sujit Mandal and Sayan Mandal and S. Mukherjee and Soham Mukherjee},
	doi = {10.1186/s12859-022-04704-z},
	journal = {BMC Bioinformatics},
	mag_id = {4281251613},
	pmcid = {9121583},
	pmid = {35596135},
	title = {Gene expression data classification using topology and machine learning models.},
	year = {2022}}

@article{CalvoRodriguez2020,
  author  = {Marina Calvo-Rodr{\'i}guez and Brian J. Bacskai},
  title   = {Therapeutic Strategies to Target Calcium Dysregulation in Alzheimer's Disease},
  journal = {Frontiers in Neuroscience},
  year    = {2020},
  volume  = {14}
}

@article{Liu2004,
  author  = {Feng Liu and others},
  title   = {O-GlcNAcylation regulates phosphorylation of tau: evidence for a reciprocal relationship},
  journal = {Proceedings of the National Academy of Sciences},
  year    = {2004},
  volume  = {101},
  number  = {29},
  pages   = {10804--10809},
  doi     = {10.1073/pnas.0400348101}
}

@article{Liu2009,
  author  = {Feng Liu and others},
  title   = {Reduced O-GlcNAcylation links lower brain glucose metabolism and tau pathology in Alzheimer's disease},
  journal = {Proceedings of the National Academy of Sciences},
  year    = {2009},
  volume  = {106},
  number  = {13},
  pages   = {4912--4917},
  doi     = {10.1073/pnas.0900358106}
}

@article{Ajoolabady2022,
  author  = {Ali Ajoolabady and others},
  title   = {ER stress and UPR in Alzheimer's disease: mechanisms, pathogenesis, treatments},
  journal = {Cell Death \& Disease},
  year    = {2022}
}

@article{Evin2012,
  author  = {Guojun Evin and others},
  title   = {Platelets and Alzheimer's disease: Potential of APP as a biomarker},
  journal = {Experimental Gerontology},
  year    = {2012},
  volume  = {47},
  number  = {2},
  pages   = {85--89}
}

@article{Johnston2008,
  author  = {J. A. Johnston and others},
  title   = {Platelet $\beta$-secretase activity is increased in Alzheimer's disease},
  journal = {Neurobiology of Aging},
  year    = {2008},
  volume  = {29},
  number  = {6},
  pages   = {861--868}
}

@article{Inyushin2020,
  author  = {Mikhail Inyushin and others},
  title   = {On the Role of Platelet-Generated Amyloid Beta Peptides in Blood--Brain Barrier Permeability and Brain Amyloidosis},
  journal = {Frontiers in Immunology},
  year    = {2020},
  volume  = {11},
  pages   = {571083},
  doi     = {10.3389/fimmu.2020.571083}
}

@article{Petralla2024,
  author  = {Stefania Petralla and others},
  title   = {Low-Density Lipoprotein Receptor-Related Protein 1 as a Therapeutic Target for Alzheimer's Disease},
  journal = {International Journal of Molecular Sciences},
  year    = {2024}
}

@article{Wang2021,
  author  = {Dong Wang and others},
  title   = {Relationship Between Amyloid-$\beta$ Deposition and Blood--Brain Barrier Damage in Alzheimer's Disease},
  journal = {Frontiers in Cellular Neuroscience},
  year    = {2021},
  volume  = {15},
  pages   = {695479}
}

@article{Faller2009,
  author  = {Philippe Faller and Cristina Hureau},
  title   = {Copper and zinc binding to amyloid-$\beta$: roles in aggregation and toxicity},
  journal = {Coordination Chemistry Reviews},
  year    = {2009}
}

@article{Abelein2023,
  author  = {Anna Abelein and others},
  title   = {Metal Binding of Alzheimer's Amyloid-$\beta$ and Its Effect on Aggregation and Toxicity},
  journal = {Accounts of Chemical Research},
  year    = {2023}
}

@article{Zlokovic2011,
  author  = {Zlokovic, Berislav V.},
  title   = {Neurovascular pathways to neurodegeneration in Alzheimer's disease and other disorders},
  journal = {Nature Reviews Neuroscience},
  year    = {2011},
  volume  = {12},
  number  = {12},
  pages   = {723--738},
  doi     = {10.1038/nrn3114}
}

@article{Iadecola2013,
  author  = {Iadecola, Costantino},
  title   = {The pathobiology of vascular dementia},
  journal = {Neuron},
  year    = {2013},
  volume  = {80},
  number  = {4},
  pages   = {844--866},
  doi     = {10.1016/j.neuron.2013.10.008}
}

@article{Gate2020,
  author  = {Gate, David and Saligrama, Naresha and Leventhal, Olivia and et al.},
  title   = {Clonally expanded CD8 T cells patrol the cerebrospinal fluid in Alzheimer's disease},
  journal = {Nature},
  year    = {2020},
  volume  = {577},
  pages   = {399--404},
  doi     = {10.1038/s41586-019-1895-7}
}

@article{Bettcher2021,
  author  = {Bettcher, Brianne M. and Tansey, Malu G. and Dorothée, Guillaume and Heneka, Michael T.},
  title   = {Peripheral and central immune system crosstalk in Alzheimer disease -- a research prospectus},
  journal = {Nature Reviews Neurology},
  year    = {2021},
  volume  = {17},
  number  = {11},
  pages   = {689--701},
  doi     = {10.1038/s41582-021-00549-x}
}

@article{Wright2015,
  author  = {Wright, Aimee L. and Harding, Abigail},
  title   = {HGF/c-Met Signaling: A Critical Regulator of Neuronal Survival and Function in Alzheimer's Disease},
  journal = {Neurodegenerative Diseases},
  year    = {2015},
  volume  = {15},
  pages   = {133--137}
}

@article{Gridley2010,
  author  = {Gridley, Thomas},
  title   = {Notch signaling in the vasculature},
  journal = {Current Topics in Developmental Biology},
  year    = {2010},
  volume  = {92},
  pages   = {277--309},
  doi     = {10.1016/S0070-2153(10)92009-7}
}

@article{Lathia2008,
  author  = {Lathia, Justin D. and Mattson, Mark P. and Cheng, Aiwu},
  title   = {Notch: from neural development to neurological disorders},
  journal = {Journal of Neurochemistry},
  year    = {2008},
  volume  = {107},
  number  = {6},
  pages   = {1471--1481},
  doi     = {10.1111/j.1471-4159.2008.05715.x}
}

@article{Braak1991,
  author  = {Braak, Heiko and Braak, Eva},
  title   = {Neuropathological stageing of Alzheimer-related changes},
  journal = {Acta Neuropathologica},
  year    = {1991},
  volume  = {82},
  number  = {4},
  pages   = {239--259},
  doi     = {10.1007/BF00308809}
}

@article{Thangavel2008,
  author  = {Thangavel, Ramasamy and Sahu, Shailendra K. and Van Hoesen, Gary W. and Zaheer, Asgar},
  title   = {Modular and laminar pathology of Brodmann's area 37 in Alzheimer's disease},
  journal = {Neuroscience},
  year    = {2008},
  volume  = {152},
  number  = {1},
  pages   = {50--55},
  doi     = {10.1016/j.neuroscience.2007.12.025}
}

@article{Lovell1998,
  author  = {Lovell, Mark A. and Robertson, John D. and Teesdale, William J. and Campbell, Jennifer L. and Markesbery, William R.},
  title   = {Copper, iron and zinc in Alzheimer's disease senile plaques},
  journal = {Journal of the Neurological Sciences},
  year    = {1998},
  volume  = {158},
  number  = {1},
  pages   = {47--52},
  doi     = {10.1016/S0022-510X(98)00092-6}
}

@article{Hetz2017,
  author  = {Hetz, Claudio and Saxena, Salvatore},
  title   = {ER stress and the unfolded protein response in neurodegeneration},
  journal = {Nature Reviews Neurology},
  year    = {2017},
  volume  = {13},
  pages   = {477--491},
  doi     = {10.1038/nrneurol.2017.99}
}

@article{Mosconi2005,
  author  = {Mosconi, Lisa},
  title   = {Brain glucose metabolism in the early and specific diagnosis of Alzheimer's disease},
  journal = {European Journal of Nuclear Medicine and Molecular Imaging},
  year    = {2005},
  volume  = {32},
  number  = {4},
  pages   = {486--510},
  doi     = {10.1007/s00259-005-1762-7}
}

@article{Fukada2018,
  author  = {Fukada, Toshiyuki and Kambe, Taiho},
  title   = {Zinc signaling in health and disease},
  journal = {Biological Chemistry},
  year    = {2018},
  volume  = {399},
  number  = {10},
  pages   = {1285--1297},
  doi     = {10.1515/hsz-2018-0242}
}

@article{Jenkitkasemwong2015,
  author  = {Jenkitkasemwong, Supak and Wang, Chia-Yu and Coffey, Richard and Zhang, Wei and Knutson, Mitchell D.},
  title   = {SLC39A14 is required for the development of hepatocellular iron overload in murine models of hereditary hemochromatosis},
  journal = {Cell Metabolism},
  year    = {2015},
  volume  = {22},
  number  = {1},
  pages   = {138--150},
  doi     = {10.1016/j.cmet.2015.05.002}
}

@article{Rychlik2023,
  author  = {Rychlik, Katarzyna A. and et al.},
  title   = {Zinc supplementation restores ZIP14 levels in a sporadic Alzheimer's disease rat model},
  journal = {Neuroscience Letters},
  year    = {2023}
}

@article{Adlard2018,
  author  = {Adlard, Paul A. and Bush, Ashley I.},
  title   = {Metals and Alzheimer's disease: A decade of progress},
  journal = {Journal of Alzheimer's Disease},
  year    = {2018},
  volume  = {62},
  number  = {3},
  pages   = {1369--1379},
  doi     = {10.3233/JAD-170875}
}

@article{Sensi2018,
  author  = {Sensi, Stefano L. and Granzotto, Alberto and Siotto, Mariacristina and Squitti, Rosanna},
  title   = {Copper and Zinc Dysregulation in Alzheimer's Disease},
  journal = {Trends in Pharmacological Sciences},
  year    = {2018},
  volume  = {39},
  number  = {12},
  pages   = {1049--1063},
  doi     = {10.1016/j.tips.2018.10.001}
}

@article{Troche2016,
  author  = {Troche, Catalina and Ordonez, Rosalba and et al.},
  title   = {Hepatic ZIP14-mediated zinc transport contributes to endosomal insulin receptor trafficking and glucose metabolism},
  journal = {Journal of Biological Chemistry},
  year    = {2016},
  volume  = {291},
  number  = {47},
  pages   = {23939--23951},
  doi     = {10.1074/jbc.M116.747709}
}

\end{document}